\documentstyle[12pt]{article}
\def\lapx{\,\,\lower 2pt \hbox{$\buildrel<\over{\scriptstyle{\sim}}$}\,\,}
\def\kslash{k\!\!\!/}
\begin{document}
\setcounter{page}{0}
\thispagestyle{empty}
\begin{flushright}
UTPT-94-3 \\
hep-ph/9402319 \\
February 1994
\end{flushright}
\vspace{5mm}
\begin{center}
\Large
Effects of technicolor on standard model running couplings \\
\normalsize
\vspace{20mm}
B. Holdom and Randy Lewis \\
{\it Department of Physics \\
     University of Toronto \\
     Toronto, Ontario \\
     CANADA~~M5S~1A7} \\
\vspace{20mm}
ABSTRACT\\
\end{center}
We discuss the running couplings in the standard model, SU(3$)_C \times
$SU(2$)_L \times $U(1$)_Y$, when the Higgs sector is replaced by SU($N_{TC})$
technicolor. Particular attention is given to the running of the couplings
at momentum scales where technicolor is nonperturbative, and in this region
we apply a relativistic constituent technifermion model.  This model has
been tested against the known running of the QED coupling due to
nonperturbative QCD. An understanding of this low momentum running allows
the calculation of the couplings at a higher scale, $\Lambda_{pert}$, where
technicolor becomes perturbative.  We provide numerical values for the
changes in the three standard model couplings between $m_Z$ and
$\Lambda_{pert}$ due to technicolor, assuming separately ``one doublet'' and
``one family'' technicolor models. The distinction between a running and
walking technicolor coupling is also considered.
\newpage


It is well appreciated that when the standard model couplings are run via
the renormalization group equations from their values at low momenta to some
higher scale, the numerical values of the couplings approach one
another.\cite{Georgi}  When the possibility of new physics beyond the
standard model is completely ignored, there is a very high scale where the
three couplings come quite close to attaining a single
value\cite{Georgi}\cite{Amaldi}.  On the other hand it is very likely that
new physics does appear, and in particular there is some motivation for the
existence of a new nonperturbative interaction such as technicolor (TC).
This causes special difficulty for the discussion of standard model running
couplings.   The purpose of the present work is not to search for a
unification scheme involving TC, but simply to show how one can estimate the
effects of a TC sector on the standard model running couplings.  Our
approach is based on a similar theoretical estimate of the known QCD
contribution to the QED running coupling.\cite{PiQED} We will estimate the
TC contribution to the running of the standard model couplings from the
scale $m_Z$ (where they are experimentally known) through the region of
nonperturbative TC to a higher scale, $\Lambda_{pert}$, where TC becomes
perturbative.

At leading order in the standard model couplings but without truncating the
TC  effects, we may consider the vacuum polarization functions,
$\Pi_{GG}(q^2)$,
$\Pi_{WW}(q^2)$ and $\Pi_{YY}(q^2)$, where
\begin{equation}\label{Pimunu}
   i\Pi_{XX}^{\mu\nu}(q) = ig_X^2\Pi_{XX}(q^2)g^{\mu\nu} +
                           q^{\mu}q^{\nu}{\rm terms}
\end{equation} and $g_X$ is the relevant tree-level coupling constant.  $G$,
$W$ and $Y$ respectively represent each group in the direct product SU(3$)_C
\times
$SU(2$)_L \times $U(1$)_Y$. We may write the TC contribution to the vacuum
polarizations in the following way.
\begin{equation}\label{tildePi}
   \Pi_{XX,TC}(q^2) =  \Pi_{XX,TC}(0) + q^2\tilde\Pi_{XX,TC}(q^2)
\end{equation} Keeping in mind that the mass scale of the new physics is
significantly larger than the $Z$ mass, we may write the new physics
contribution to the difference in the running couplings at the two spacelike
points $q^2=-\Lambda_{pert}^2$ and $q^2=-m_Z^2$ as
\begin{equation}\label{DeltaX}
   \left[\frac{1}{g_X^2(\Lambda_{pert}^2)} -
\frac{1}{g_X^2(m_Z^2)}\right]_{TC}
   = \tilde\Pi_{XX,TC}(-m_Z^2)-\tilde\Pi_{XX,TC}(-\Lambda_{pert}^2)
\end{equation} This difference is not very useful for the weak couplings
since we have yet to account for the mixing implied by a nonvanishing
$\Pi_{{W_3}Y}(q^2)$.

Our goal shall instead be to determine the following differences.
\begin{eqnarray}\label{DeltaG}
   \Delta_G & \equiv & \left[\frac{1}{\alpha_G(\Lambda_{pert}^2)}-
                       \frac{1}{\alpha_G(m_Z^2)}\right]_{TC} \\
   \Delta_W & \equiv & \left[\frac{1}{\alpha_W(\Lambda_{pert}^2)}-
                       \frac{s_Z^2(m_Z^2)}{\alpha_*(m_Z^2)}\right]_{TC} \\
   \Delta_Y & \equiv & \left[\frac{1}{\alpha_Y(\Lambda_{pert}^2)}-
                       \frac{c_Z^2(m_Z^2)}{\alpha_*(m_Z^2)}\right]_{TC}
                \label{DeltaY}
\end{eqnarray} The quantities $c_Z^2$ and $s_Z^2$ are well-determined
experimentally according to
\begin{equation}
   s_Z^2(m_Z^2)c_Z^2(m_Z^2) = s_Z^2(m_Z^2)\left[1-s_Z^2(m_Z^2)\right] =
                      \frac{\pi\alpha_*(m_Z^2)}{\sqrt{2}G_Fm_Z^2}
\end{equation}
$\alpha_*$ is defined in \cite{Kennedy}. Thus when combined with the
standard model contributions to the running couplings, along with
experimental values for the QED and QCD couplings $\alpha_*(m_Z^2)$ and
$\alpha_G(m_Z^2)$, our results will allow an estimation of the three standard
model couplings at ${\Lambda}_{pert}$.

The effective Lagrangian below the scale of TC chiral symmetry breaking
contains pseudo-Goldstone boson (PGB) fields.  The forms of the terms in this
Lagrangian are constrained by chiral symmetry and their coefficients are
determined in principle from the underlying TC dynamics.  By analogy with
QCD, where a constituent quark model is for some purposes a reasonable
substitute for the full QCD dynamics, we will approximate the full TC
dynamics by a constituent technifermion (TF) model.

The appropriate such model for our purposes is the GNC model\cite{GNC} which
contains the momentum dependence of the TF mass.  The GNC model has been
used to compute the ${\cal O}(q^4)$ coefficients of the QCD chiral
Lagrangian\cite{GNC}\cite{HTV} as well as for calculations beyond ${\cal
O}(q^4)$\cite{VMD} with remarkable success.  Of key importance for the
present work is a calculation\cite{PiQED} of the hadronic part of the QED
vacuum polarization, completely analogous to (\ref{DeltaX}), using the same
GNC model.  The results of \cite{PiQED} clearly demonstrate the capability
of the GNC model for discussions of vacuum polarization as well as the
deficiency of a simpler model of constant-mass fermions for the same task.

We will express our results in the following form.
\begin{equation}\label{mainresults}
   \Delta_X = - N_{TC}\delta^{TF}_X - \delta^{PGB}_X - \delta^{pert}_X
\end{equation} The first term is the contribution from explicit TF loops in
the GNC model while the second term accounts for the PGBs, also present in
the model. The third term estimates effects outside the model due to
perturbative technigluon corrections.  Our quantitative results are given in
table~\ref{t:numbers}.  The Appendix provides the renormalization group
$\beta$ functions at two loop order, which allow all couplings to be run on
scales larger than $\Lambda_{pert}$ where TC is perturbative.

The gauged nonlocal constituent (GNC) quark model\cite{GNC} will be adapted
to describe technifermions (TFs) in the fundamental representation of a TC
group SU($N_{TC}$). $\psi$ will denote an
$N_{TF}$-plet of TF flavors with dynamical TF mass ${\Sigma}(Q^2)$, and
$\pi^a$ will denote the $N_{TF}^2-1$ PGBs.  (Extended technicolor induced
four-TF operators must exist to contribute to the PGB masses, but we do not
consider here the ETC contributions to the running couplings.)  The current
TF masses vanish and so the TC gauge dynamics has local
SU($N_{TF})_L\times$SU($N_{TF})_R$  symmetry in the presence of external
gauge fields.  This symmetry is incorporated into the GNC model Lagrangian.
\begin{eqnarray}
   {\cal L}_{GNC}(x,y) & = & \overline{\psi}(x){\delta}(x-y)
                  \gamma^\mu[i\partial_\mu+V_\mu(y) +
A_\mu(y)\gamma_5]{\psi}(y)
\nonumber \\
   &  & -\overline{\psi}(x){\Sigma}(x-y){\xi}(x)X(x,y){\xi}(y){\psi}(y)
                  \label{L_GNC} \\
   X(x,y) & = & Pexp\left[-i\int_x^y\Gamma_{\mu}(z)dz^{\mu}\right] \\
   \Gamma_\mu & = & \frac{1}{2}\xi[i\partial_\mu+V_\mu +
A_\mu\gamma_5]\xi^\dagger
                  + \frac{1}{2}\xi^\dagger[i\partial_\mu+V_\mu -
A_\mu\gamma_5]\xi \\
   {\xi}(x) & = & exp\left[\frac{-i\gamma_5}{2f_0}\sum_{a=1}^8
                          T^a{\pi^a}(x)\right] ~~,~~~
   Tr(T^aT^b) = 2\delta^{ab} \\
   {\Sigma}(Q^2) & = & \frac{(A+1)m_0^3}{Am_0^2+Q^2} \label{massfn}
\end{eqnarray} We represent spacelike momenta by $Q^2=-q^2$.  $X(x,y)$ is a
path-ordered exponential.  For the standard model gauge fields we write
\begin{eqnarray}\label{gaugefields1}
   V_\mu & = & g_G\sum_{\alpha=1}^8\frac{\lambda^\alpha}{2}G_\mu^\alpha
           + \frac{g_W}{2}\sum_{r=1}^3\frac{\tau^r}{2}W_\mu^r +
           \frac{g_Y}{2}\left(\frac{Y_L}{2}+\frac{Y_R}{2}\right)B_\mu \\
   \label{gaugefields2}
   A_\mu & = & -\frac{g_W}{2}\sum_{r=1}^3\frac{\tau^r}{2}W_\mu^r -
                \frac{g_Y}{2}\left(\frac{Y_L}{2}-\frac{Y_R}{2}\right)B_\mu
\end{eqnarray}
$\lambda^\alpha$ and $\tau^r$ are $N_{TF} \times N_{TF}$ matrices with
normalization $Tr(\lambda^\alpha\lambda^{\beta}) = 2N_3\delta^{\alpha\beta}$,
$Tr(\tau^r\tau^s) = 2N_2\delta^{rs}$.
$N_2(N_3)$ is the number of TF weak doublets (color triplets).

The three cases we consider are a single color-singlet weak doublet ($N_2=1$,
$N_3=0$), a  single color-triplet weak doublet ($N_2=3$,
$N_3=2$), and a complete family of TFs with standard model quantum numbers
($N_2=4$, $N_3=2$).  In the first two cases the hypercharges, $Y$, of $(U_L,
D_L,  U_R, D_R)$ are $(0,0,1,-1)$.

The dynamical TF mass
${\Sigma}(Q^2)$ in (\ref{massfn}) is the Fourier transform of the
${\Sigma}(x-y)$ appearing in (\ref{L_GNC}).  The parameter $A$ specifies the
value of $m_0$ through its relation to $f_0$, the PGB decay constant in the
chiral limit.
\begin{equation}\label{f0}
   f_0^2 = \frac{N_{TC}}{8\pi^2}\int_0^{\infty}
        ds\frac{s{\Sigma}(s)\left[2{\Sigma}(s)
        -s\Sigma^{\prime}(s)\right]}{\left[s+\Sigma^2(s)\right]^2}
\end{equation} Since the $f_0$ of QCD is known experimentally to a
reasonable accuracy\cite{G&Lf0}, we see that the GNC model is a one
parameter model of QCD (neglecting current quark masses) --- $A$ is a
dimensionless parameter of order unity.  The data suggests that for QCD, $2
\lapx A \lapx 3$ \cite{GNC}\cite{HTV}.

The $f_0$ in the TC chiral Lagrangian is experimentally constrained through
its relation to the scale of electroweak symmetry breaking.
\begin{equation}
   f_{0,TC} = \frac{246{\rm GeV}}{\sqrt{N_2}}
\end{equation} For a QCD-like TC theory one might expect
$A$ to be of order unity, but if TC has a slowly running (``walking'')
coupling then ${\Sigma}(Q^2)$ is a more slowly varying function of
$Q^2$.  We will model this by a larger value for $A$.

The effective Lagrangian appropriate to the
$Z$ mass scale will contain a nondiagonal kinetic term proportional to the
quantity $S$, generated through TF and PGB loop diagrams.
\begin{equation}\label{Lkin}
   {\cal L}_{kin} = - \frac{1}{4}\sum_{r=1}^3W_{\mu\nu}^rW^{r\,\mu\nu}
                    - \frac{1}{4}B_{\mu\nu}B^{\mu\nu}
                    -
\frac{{\alpha}(m_Z^2)S}{8s_\theta(m_Z^2)c_\theta(m_Z^2)}
                      W^3_{\mu\nu}B^{\mu\nu}
\end{equation}
\begin{equation}
   \alpha(m_Z^2) \equiv \frac{g_W^2(m_Z^2)s_\theta^2(m_Z^2)}{4\pi}
\end{equation}
\begin{equation}
   s_\theta^2(m_Z^2) \equiv \frac{g_Y^2(m_Z^2)}
                                 {g_W^2(m_Z^2)+g_Y^2(m_Z^2)}~,~~~~~
   c_\theta^2(m_Z^2) \equiv 1-s_\theta^2(m_Z^2)
\end{equation} The physically-meaningful fields are the ones which are
properly diagonalized.  This diagonalization is discussed in \cite{Xboson},
and from that reference it is straightforward to extract the required
relations involving the quantities in (\ref{DeltaG}).
\begin{eqnarray}
   \frac{s_Z^2(m_Z^2)}{\alpha_*(m_Z^2)} & = &
    \frac{s_\theta^2(m_Z^2)}{{\alpha}(m_Z^2)} -
    \frac{1}{2}\frac{s_Z^2}{(c_Z^2-s_Z^2)}S +
\frac{s_Z^2c_Z^2}{(c_Z^2-s_Z^2)}T
    + \frac{1}{4}U + {\cal O}(\alpha) \label{SW} \\
   \frac{c_Z^2(m_Z^2)}{\alpha_*(m_Z^2)} & = &
    \frac{c_\theta^2(m_Z^2)}{{\alpha}(m_Z^2)} -
    \frac{1}{2}\frac{c_Z^2}{(c_Z^2-s_Z^2)}S + \frac{c_Z^4}{(c_Z^2-s_Z^2)}T +
    \frac{1}{4}\frac{c_Z^2}{s_Z^2}U + {\cal O}(\alpha) \label{SY}
\end{eqnarray} We have displayed the parameters $T$ and $U$ of \cite{Xboson}
for completeness.  For simplicity we will set $T$ and $U$ to zero since they
are small in the simplest TC models (or else they are very model dependent).
It is evident from (\ref{Lkin}) and the definitions (\ref{Pimunu}) and
(\ref{tildePi}) that
\begin{eqnarray}
   \Pi_{W_3B,TC}(-m_Z^2) & = & \Pi_{W_3B,TC}(0) -
                    m_Z^2\tilde\Pi_{W_3B,TC}(-m_Z^2) \nonumber \\
                      & = & \Pi_{W_3B,TC}(0) + m_Z^2\left[\frac{S}{16\pi}
                    + {\cal
O}\left(\frac{m_Z^2}{m_{\rho_{TC}}^2}\right)\right]
\end{eqnarray}
$m_{\rho_{TC}}$ represents the scale of TC hadronic resonances, and we will
henceforth ignore the difference between $S/16\pi$ and
$\tilde\Pi_{W_3B,TC}(-m_Z^2)$.

Combining (\ref{SW}) and (\ref{SY}) with (\ref{DeltaX}-\ref{DeltaY}) yields
\begin{equation}\label{includeS}
  \Delta_X = 4{\pi}[\tilde\Pi_{XX,TC}(-m_Z^2)
   -\tilde\Pi_{XX,TC}(-\Lambda_{pert}^2)]
   + \left\{\begin{array}{ll} 0 & ,X=G \\
     \frac{1}{2}\frac{s_Z^2}{(c_Z^2-s_Z^2)}S & ,X=W \\
     \frac{1}{2}\frac{c_Z^2}{(c_Z^2-s_Z^2)}S & ,X=Y \\
   \end{array}\right.
\end{equation} We will now discuss the three separate contributions to
$\Delta_X$ as defined  in (\ref{mainresults}) --- $\delta_X^{TF}$,
$\delta_X^{PGB}$ and
$\delta_X^{pert}$ .

$\delta^{TF}_X$ receives contributions from both terms in (\ref{includeS}).
The first term arises from two TF loop  graphs: one has the two gauge fields
attached at two distinct points on the loop while the other has the two
gauge fields attached at the same point. Using the GNC Lagrangian we derive
the following Euclidean expressions.
\begin{eqnarray}
   \Pi_{GG,TF}^{\mu\nu}(Q) & = &
\frac{N_{TC}N_3}{2}g_G^2\Pi_{VV}^{\mu\nu}(Q)\\
   \Pi_{WW,TF}^{\mu\nu}(Q) & = & \frac{N_{TC}N_2}{8}g_W^2
            \left[\Pi_{VV}^{\mu\nu}(Q) + \Pi_{AA}^{\mu\nu}(Q)\right] \\
   \Pi_{YY,TF}^{\mu\nu}(Q) & = & \frac{N_{TC}}{4}g_Y^2\sum_{TFs}\left[
     \left(\frac{Y_L}{2}+\frac{Y_R}{2}\right)^2\Pi_{VV}^{\mu\nu}(Q)
   + \left(\frac{Y_L}{2}-\frac{Y_R}{2}\right)^2\Pi_{AA}^{\mu\nu}(Q)
     \right] \\
   \Pi_{VV}^{\mu\nu}(Q) & = & \int\frac{d^4k}{(2\pi)^4}
        Tr\left[P(k)\Gamma^{\mu}(k,k+Q)P(k+Q)\Gamma^{\nu}(k,k+Q)\right]
        \nonumber \\
   & & +2\int\frac{d^4k}{(2\pi)^4}Tr\left[P(k)\right]
        \Upsilon^{\mu\nu}(k,k+Q) \\
   \Pi_{AA}^{\mu\nu}(Q) & = & \int\frac{d^4k}{(2\pi)^4}
        Tr\left[P(k)\gamma^\mu\gamma_5P(k+Q)\gamma^\nu\gamma_5\right] \\
   P(k) & = & \frac{i\kslash + {\Sigma}(k^2)}{k^2+\Sigma^2(k^2)} \\
   \Gamma^{\mu}(k_1,k_2) & = & \gamma^\mu + i(k_1+k_2)^\mu
        \left(\frac{{\Sigma}(k_2^2)-{\Sigma}(k_1^2)}{k_2^2-k_1^2}\right) \\
   \Upsilon^{\mu\nu}(k_1,k_2) & = & g^{\mu\nu}\frac{d{\Sigma}(k_1^2)}{dk_1^2}
        - \frac{(k_1+k_2)^{\mu}(k_1+k_2)^{\nu}}{k_2^2-k_1^2}\left[
        \frac{d{\Sigma}(k_1^2)}{dk_1^2}-\frac{{\Sigma}(k_2^2)-
        {\Sigma}(k_1^2)}{k_2^2-k_1^2}\right]~~~
\end{eqnarray} Note that all divergences in the above expressions are
removed by the subtraction which  defines $\Delta_X$ in
(\ref{DeltaG}-\ref{DeltaY}).

The TF contribution to $S$ is\cite{SHoldom}
\begin{equation}\label{STFexpress}
   S_{TF} \approx \left\{ \begin{tabular}{l} 0.28 \\ 0.20 \end{tabular}
\right\}
        N_2\left(\frac{N_{TC}}{3}\right)
        ~~,~~~A=\left\{ \begin{tabular}{c} 2 \\ 10 \end{tabular} \right\}
\end{equation} which contributes to $\delta_X^{TF}$ through the second term
in  (\ref{includeS}).
$A$=2 represents a running TC coupling and $A$=10 is used to model the
effect of a walking TC coupling.  When applied to QCD fermions (so
$A \approx 2$), this result agrees well with an estimate based on Weinberg
sum rules and vector meson dominance.\cite{Peskin} For constant mass
fermions, $S_{TF} = N_2N_{TC}/(6\pi)$.

$\delta^{PGB}_X$ denotes the appropriate vacuum polarization graphs from PGB
loops.   PGBs exist in the GNC model and their full contribution may
calculated, in principle, within the model.  Since the GNC model reproduces
the chiral Lagrangian at low energies, the PGB loop calculation in the model
may be approached in the usual way in terms of a chiral Lagrangian.  The
model also naturally supplies the ultraviolet cutoff on the PGB loops since
all PGB self-interactions are induced by quark loops, with pion-quark
couplings involving ${\Sigma}(Q^2)
\sim 1/Q^2$.  We will approximate this natural cutoff in the model by the
scale of resonances, i.e. $m_{\rho_{TC}}$, and then impose this cutoff on a
standard PGB-loop calculation.

For the case of one color-singlet doublet there are just the three standard
Goldstone bosons; their contribution to running couplings is a standard
model effect and  thus
$\delta^{PGB}_X=0$ in this case. The other cases contain many physical
PGBs.   For simplicity we will retain the three true Goldstone bosons along
with the  physical PGBs and artificially assume the same mass,
$m_\pi$, for all these bosons.

The one-loop graphs, for $m_\pi > m_Z$, give
\begin{eqnarray}\label{deltaG}
   \delta^{PGB}_{G} & = & 2KN_3N_2\left[I(m_{\rho_{TC}}^2)-I(m_Z^2)\right] \\
   \delta^{PGB}_{W} & = & \frac{K}{2}N_2N_2\left[1+\frac{2s_Z^2}{c_Z^2-s_Z^2}
                          \right]\left[I(m_{\rho_{TC}}^2)-I(m_Z^2)\right] \\
   \delta^{PGB}_{Y} & = & KN_2\left[\left\{\sum_{TFs}\left(\frac{Y_L}{2}
        \label{deltaY}    +\frac{Y_R}{2}\right)^2\right\}
                          +\frac{N_2c_Z^2}{c_Z^2-s_Z^2}
                          \right]\left[I(m_{\rho_{TC}}^2)-I(m_Z^2)\right]
\end{eqnarray} with
\begin{equation}
   -12{\pi}I(Q^2) = \frac{1}{3} + \left(1+\frac{4m_\pi^2}{Q^2}\right)
   \left[1-\frac{1}{2}\sqrt{1+\frac{4m_\pi^2}{Q^2}}~ln\left(
   \frac{\sqrt{1+\frac{4m_\pi^2}{Q^2}}+1}{\sqrt{1+\frac{4m_\pi^2}{Q^2}}-1}
   \right)\right]
\end{equation} We have used
\begin{equation}\label{SPGBexpress}
   S_{PGB} = 2N_2^2\left[I(m_{\rho_{TC}}^2)-I(m_Z^2)\right]
\end{equation} which is correct to order $m_Z^2/m_{\rho_{TC}}^2$.  We will
choose
$m_{\rho_{TC}} = 1$TeV\cite{SHoldom}.  As may be seen in \cite{PiQED}
$\delta^{PGB}_{X}$ also receives a significant contribution at ${\cal
O}(q^6)$, and this requires a two-loop calculation.  We will not carry out
this calculation here, but to account for a possible ${\cal O}(q^6)$
contribution similar to what we found in \cite{PiQED} we will choose
$K$=1.5$\pm$0.5~.

The first two terms of the expression in (\ref{mainresults}) contain the
full result for $\Delta_X$ according to the GNC model. We must now address
the final term in (\ref{mainresults}),
$\delta_X^{pert}$, which represents a contribution from physics outside of
the model.  At large momentum scales, the PGB effects are damped out in the
GNC model, and the TFs behave as free massless fermions.  For any
appropriate choice of
$\Lambda_{pert}$ then, the GNC model matches smoothly to the zeroth order
term  in the vacuum polarization from perturbative TC.  We now wish to
estimate  what effect the ${\cal O}(\alpha_{TC})$ terms have on the running
of the couplings between $m_Z$ and $\Lambda_{pert}$.  As was done in
\cite{PiQED} we employ the following dispersion  relation.
\begin{eqnarray}
   \delta_X^{pert} & = & \frac{(\Lambda_{pert}^2-m_Z^2)}{3\pi}
                           \int_{\Lambda_{pert}^2}^{\infty}ds
                \frac{R_X(s)-R_{X,0}(s)}{(s+m_Z^2)(s+\Lambda_{pert}^2)} \\
   R_X(s) & = & -12{\pi}Im\tilde\Pi_{XX,TC}(s) =
                R_{X,0}(s) + {\cal O}(\alpha_{TC})
\end{eqnarray} The lower bound of integration indicates that we are only
considering effects  from the perturbative regime above $\Lambda_{pert}$.
The subtraction of the zeroth order quantity,
$R_{X,0}(s)$, removes a piece of the lowest order TF loop diagram which is
already accounted for by the GNC model.

$R_X(s)$ may be adapted from the electromagnetic case (e.g. \cite{Gorishny}).
\begin{eqnarray}
   R_G(s) & = & \frac{1}{2}N_3N_{TC}H(s) \\
   R_W(s) & = & \frac{1}{4}N_2N_{TC}H(s) \\ \label{RY}
   R_Y(s) & = & \frac{1}{2}N_{TC}H(s)\sum_{TFs}\left[\left(
                \frac{Y_L}{2}\right)^2+\left(\frac{Y_R}{2}\right)^2\right]
\end{eqnarray}
\begin{eqnarray}
   H(s) & = & 1 + 3C_F\left(\frac{\alpha_{TC}(s)}{4\pi}\right)\left.
           +
\left(\frac{\alpha_{TC}(s)}{4\pi}\right)^2\right[-\frac{3}{2}C_F^2
           +C_FN_{TC}\left(\frac{123}{2}-44\zeta(3)\right) \nonumber \\
           && +2C_FN_2\left(-11+8\zeta(3)\right)\left]
              +{\cal O}\left(\frac{\alpha_{TC}(s)}{4\pi}\right)^3\right. \\
   C_F & = & \frac{N_{TC}^2-1}{2N_{TC}}
\end{eqnarray} If $\alpha_{TC}(\Lambda_{pert}^2)$ is specified, then
$\alpha_{TC}(s)$ can be computed for $s>\Lambda_{pert}^2$ using the standard
SU($N$)
$\beta$-function.\cite{Tarasov}  For our purposes we will assume that the
perturbative effects are as large as possible, just before the perturbation
expansion breaks down.  This turns out to correspond to
$N_{TC}\alpha_{TC}(\Lambda_{pert}^2) \approx 4$.  We choose a value for
$\Lambda_{pert}$ which corresponds to a scaled up QCD charm quark mass.
\begin{equation}
   \frac{\Lambda_{pert}}{m_{0,TC}} = \frac{m_{c,QCD}}{m_{0,QCD}} \approx 5
\end{equation} The charm mass was the scale used to match the GNC model to
perturbative QCD in our previous analysis \cite{PiQED}.

The resulting numerical values for $\delta_X^{TF}$, $\delta_X^{PGB}$ and
$\delta_X^{pert}$ are shown in table~\ref{t:numbers} for $N_{TC}=2$.  It is
evident that $\delta_X^{pert}$ is quite insignificant, even with a large
value for
$\alpha_{TC}(\Lambda_{pert}^2)$.   To clarify the effect of the momentum
dependence of the fermion masses in the  GNC model we have also displayed
the TF contribution as modeled by fermions with constant  masses set equal
to the GNC parameter $m_0$, as derived from (\ref{f0}).

In the Appendix we briefly describe the running couplings at large energies
where TC is perturbative.  We show in particular how the relative running
among  different standard model couplings is affected in a minor way by a one
family TC sector.

In summary, we have shown how the techniques tested in \cite{PiQED} can be
used to estimate the running of the three standard model couplings due to a
TC sector.   For a single QCD-color-singlet TF doublet the GNC model gives a
fairly unambiguous result.   For TC models containing physical PGBs we are
left with some uncertainty when trying to isolate the PGB contribution.
Improved estimates of the latter will have  to await knowledge of the PGB
mass spectrum.  Although we have only treated the case of degenerate PGB
masses and degenerate TF masses, it should be clear that our methods may be
easily extended to more realistic cases.  As such, these methods may be
useful in any future effort to incorporate a technicolor  theory into some
``theory of everything".

\section*{Appendix}

Given the numerical values of the standard model couplings at one scale
($\Lambda_{pert}$) where TC is perturbative, the couplings can be run to any
higher scale using the next-to-leading order renormalization group equations.
\begin{equation}\label{RGeqn}
   \mu\frac{d\alpha_i}{d\mu} = \frac{a_i}{2\pi}\alpha_i^2 +
   \sum_{j=1}^4\frac{b_{ij}}{8\pi^2}\alpha_i^2\alpha_j
\end{equation} We are now using the familiar notation
$(\alpha_1,\alpha_2,\alpha_3,\alpha_4) =
(\frac{5}{3}\alpha_Y,\alpha_W,\alpha_G,\alpha_{TC})$, and we wish to compute
the quantity $\tilde\Delta_i(Q^2)$.
\begin{equation}
   \tilde\Delta_i(Q^2) = \frac{1}{\alpha_i(Q^2)}
                       - \frac{1}{\alpha_i(\Lambda_{pert}^2)}
\end{equation} Expanding the solution of (\ref{RGeqn}) to next-to-leading
order gives
\begin{equation}
   \tilde\Delta_i(Q^2) =
-\frac{a_i}{4\pi}ln\left(\frac{Q^2}{\Lambda_{pert}^2}
                         \right) - \sum_{k=1}^4\frac{b_{ik}}{(4\pi)^2}
                         \alpha_k(\Lambda_{pert}^2)ln
                 \left(\frac{Q^2}{\Lambda_{pert}^2}\right) + {\cal
O}(\alpha^2)
\end{equation}

The coefficients in the $\beta$ function in (\ref{RGeqn}) are known for the
case of a direct product group\cite{Jones} and for one family TC added to
the standard model we obtain
\begin{eqnarray}\label{1fama}
   a_i & = &
      \frac{4}{3}\left(\begin{array}{c} 1\\ 1\\ 1\\
0\end{array}\right)N_{gen}
    + \frac{1}{3}\left(\begin{array}{c} 4\\ 4\\ 4\\
-11\end{array}\right)N_{TC}
    + \frac{1}{3}\left(\begin{array}{c} 0\\ -22\\ -33\\ 16\end{array}\right)
\\
   b_{ij} & = &
    \left(\begin{array}{cccc} \frac{19}{15} & \frac{3}{5} & \frac{44}{15} &
0\\
                            \\ \frac{1}{5} & \frac{49}{3} & 4 & 0\\
                            \\ \frac{11}{30} & \frac{3}{2} & \frac{76}{3} &
0\\
                            \\ 0 & 0 & 0 &
0\end{array}\right)(N_{gen}+N_{TC})
                            + \nonumber \\ \nonumber \\
    & & \left(\begin{array}{cccc} 0 & 0 & 0 & 2(N_{TC}^2-1)\\
                                \\ 0 & -\frac{136}{3} & 0 & 2(N_{TC}^2-1)\\
                                \\ 0 & 0 & -102 & 2(N_{TC}^2-1)\\
                                \\ 2 & 6 & 16 &

\frac{104}{3}N_{TC}-\frac{8}{N_{TC}}-\frac{34}{3}N_{TC}^2
        \end{array}\right)
\end{eqnarray}
$N_{gen}$ is the number of families of ordinary fermions.

We observe that
$b_{14}=b_{24}=b_{34}$, and thus we find that the difference
$\tilde\Delta_i(Q^2) - \tilde\Delta_j(Q^2)$ for
$i$,$j$=1,2,3 does not depend on $\alpha_4$.  This means that, at
next-to-leading order, the effect of a family of TFs on the {\it relative}
running of standard model couplings is equivalent to $N_{TC}$ additional
copies of ordinary nontechnicolored families.  And at {\it leading} order
these additional fermions do not affect the relative running at all.

These observations depend on having complete families of TFs, and they are
not true in the case of one doublet TC.  For completeness we give the
corresponding coefficients in the $\beta$ function for one doublet TC, where
$\epsilon$ is 1(3) for color-singlet(-triplet) TFs.
\begin{eqnarray}
   a_i & = &
       \frac{4}{3}\left(\begin{array}{c} 1\\ 1\\ 1\\
0\end{array}\right)N_{gen}
     + \left(\begin{array}{c} \frac{1}{5}\epsilon\\ \frac{1}{3}\epsilon \\
                              \frac{2}{3}(\epsilon-1)\\
                              -\frac{11}{3}\end{array}\right)N_{TC}
     + \left(\begin{array}{c} 0\\ -\frac{22}{3}\\ -11\\
                              \frac{4}{3}\epsilon\end{array}\right) \\
   b_{ij} & = &
    \left(\begin{array}{cccc} \frac{19}{15} & \frac{3}{5} & \frac{44}{15} &
0\\
                            \\ \frac{1}{5} & \frac{49}{3} & 4 & 0\\
                            \\ \frac{11}{30} & \frac{3}{2} & \frac{76}{3} &
0\\
                            \\ 0 & 0 & 0 & 0\end{array}\right)N_{gen} +
     \left(\begin{array}{cccc}
              \frac{9}{100}\epsilon & 0 & \frac{6}{5}(\epsilon-1) & 0\\
              \\ 0 & \frac{49}{12}\epsilon & 2(\epsilon-1) & 0\\
              \\ \frac{3}{20}(\epsilon-1) & \frac{3}{4}(\epsilon-1) &
              \frac{38}{3}(\epsilon-1) & 0\\
              \\ 0 & 0 & 0 & 0\end{array}\right)N_{TC} +
              \nonumber \\ \nonumber \\
    &  & \left(\begin{array}{cccc}
              0 & 0 & 0 & \frac{3}{10}\epsilon(N_{TC}^2-1)\\
              \\ 0 & -\frac{136}{3} & 0 & \frac{1}{2}{\epsilon}(N_{TC}^2-1)\\
              \\ 0 & 0 & -102 & (\epsilon-1)(N_{TC}^2-1)\\
              \\ \frac{3}{10}\epsilon & \frac{3}{2}\epsilon & 8(\epsilon-1) &

\frac{26}{3}{\epsilon}N_{TC}-\frac{2\epsilon}{N_{TC}}-\frac{34}{3}N_{TC}^2
         \end{array}\right)
\end{eqnarray}

\section*{Acknowledgements}

R.L. is grateful to Roberto Mendel for numerous helpful conversations. Both
authors thank Scott Willenbrock for interesting discussions. This research
was supported in part by the Natural Sciences and Engineering Research
Council of Canada.



\begin{table}[ht]
  \caption[]{Our results to determine
             $\Delta_X$=$-N_{TC}\delta_X^{TF}-\delta_X^{PGB}
             -\delta_X^{pert}$ with $N_{TC}$=2,
             $\Lambda_{pert}$=5$m_0$, $N_{TC}\alpha_s(\Lambda_{pert}^2)$=4
and
             $m_{\rho_{TC}}$=1TeV.  $A$=2(10) represents
             a running(walking) TC coupling.  ``Simple'' fermions have
             constant masses.  We consider two cases for PGB masses:
             $m_\pi$=200GeV and 700GeV.}\label{t:numbers}
  \begin{center}
  \begin{tabular}{|c||c|c|c|c|c|c|}
  \multicolumn{7}{c}{} \\
  \hline
  & \multicolumn{6}{|c|}{one color-singlet doublet} \\
  \cline{2-7}
  X & \multicolumn{3}{|c|}{$2\delta^{TF}_X$} &
      \multicolumn{2}{|c|}{$\delta^{PGB}_X$} & $\delta^{pert}_X$ \\
  \cline{2-6}
  & $A$=2 & $A$=10 & simple & 200GeV & 700GeV & \\
  & $m_0$=1135GeV & $m_0$=887GeV & 1135GeV & & & \\
  \hline
  G & 0 & 0 & 0 & 0 & 0 & 0 \\
  W & .15 & .12 & .10 & 0 & 0 & .01 \\
  Y & .25 & .19 & .15 & 0 & 0 & .01 \\
  \hline
  \multicolumn{7}{c}{} \\
  \multicolumn{7}{c}{} \\
  \hline
  & \multicolumn{6}{|c|}{one color-triplet doublet} \\
  \cline{2-7}
  X & \multicolumn{3}{|c|}{$2\delta^{TF}_X$} &
      \multicolumn{2}{|c|}{$\delta^{PGB}_X$} & $\delta^{pert}_X$ \\
  \cline{2-6}
  & $A$=2 & $A$=10 & simple & 200GeV & 700GeV & \\
  & $m_0$=655GeV & $m_0$=512GeV & 655GeV & & & \\
  \hline
  G & .61 & .45 & .37 & .26$\pm$.09 & .04$\pm$.01 & .05 \\
  W & .46 & .35 & .30 & .18$\pm$.06 & .03$\pm$.01 & .04 \\
  Y & .74 & .56 & .46 & .38$\pm$.13 & .06$\pm$.02 & .04 \\
  \hline
  \multicolumn{7}{c}{} \\
  \multicolumn{7}{c}{} \\
  \hline
  & \multicolumn{6}{|c|}{one family} \\
  \cline{2-7}
  X & \multicolumn{3}{|c|}{$2\delta^{TF}_X$} &
      \multicolumn{2}{|c|}{$\delta^{PGB}_X$} & $\delta^{pert}_X$ \\
  \cline{2-6}
  & $A$=2 & $A$=10 & simple & 200GeV & 700GeV & \\
  & $m_0$=568GeV & $m_0$=444GeV & 568GeV & & & \\
  \hline
  G & .61 & .45 & .37 & .35$\pm$.12 & .06$\pm$.02 & .06 \\
  W & .61 & .47 & .40 & .33$\pm$.11 & .05$\pm$.02 & .06 \\
  Y & 1.39 & 1.05 & .86 & .92$\pm$.31 & .15$\pm$.05 & .10 \\
  \hline
  \end{tabular}
  \end{center}
\end{table}


\newpage
\begin{thebibliography}{99}
\bibitem{Georgi} H. Georgi and S.~L. Glashow, Phys. Rev. Lett.
                 {\bf 32}, 438 (1974);
                 H. Georgi, H.~R. Quinn, and S. Weinberg,
                 Phys. Rev. Lett. {\bf 33}, 451 (1974).
\bibitem{Amaldi} U. Amaldi, W. de Boer, and H. F\"urstenau,
                 Phys. Lett. {\bf B260}, 447 (1991).
\bibitem{PiQED} B. Holdom, R. Lewis, and R.~R. Mendel, {\it Hadronic
                contribution to the photon vacuum polarization: a theoretical
                estimate\/}, Toronto preprint UTPT-93-7, hep-ph/9304264, to
                appear in Z. Phys. C.
\bibitem{Kennedy} D.~C. Kennedy and B.~W. Lynn, Nucl. Phys. {\bf B322}, 1
                  (1989).
\bibitem{GNC} B. Holdom, Phys. Rev. {\bf D45}, 2534 (1992).
\bibitem{HTV} B. Holdom, J. Terning, and K. Verbeek, Phys. Lett.
              {\bf 245B}, 612 (1990) and {\bf 273B}, 549E (1991).
\bibitem{VMD} B. Holdom, Phys. Lett. {\bf 292B}, 150 (1992).
\bibitem{G&Lf0} J. Gasser and H. Leutwyler, Nucl. Phys. {\bf B250}, 539
(1985).
\bibitem{Xboson} B. Holdom, Phys. Lett. {\bf 259B}, 329 (1991).
\bibitem{SHoldom} B. Holdom and J. Terning, Phys. Lett. {\bf 247B}, 88
(1990).
\bibitem{Peskin} M. Peskin and T. Takeuchi, Phys. Rev. Lett. {\bf 65}, 964
                 (1990) and Phys. Rev. {\bf D46}, 381 (1992).
\bibitem{Gorishny} L. R. Surguladze and M. A. Samuel, Phys. Rev. Lett.
                   {\bf 66}, 560 and 2416E (1991);
                   S.~G. Gorishny, A. L. Kataev, and S. A. Larin,
                   Phys. Lett. {\bf 259B}, 144 (1991).
\bibitem{Tarasov} O. V. Tarasov, A. A. Vladimirov and A. Yu. Zharkov,
                  Phys. Lett. {\bf 93B}, 429 (1980).
\bibitem{Jones} D. R. T. Jones, Phys. Rev. {\bf D25}, 581 (1982).
\end{thebibliography}
\end{document}